\documentclass[aps,twocolumn,superscriptaddress]{revtex4-2} 
\usepackage{color}
\usepackage{amsmath}
\usepackage{graphicx}
\usepackage{bm}
\def\ket#1{|{#1}\rangle}
\def\bra#1{\langle{#1}|}
\def\braket#1{\langle{#1}\rangle}
\usepackage{colortbl}
\usepackage{comment}
\usepackage{graphicx}
\usepackage[]{natbib}
\usepackage{hyperref}
\usepackage{url}
\usepackage{here}

\begin{document}
\title{Phonon-Mediated Spin Dynamics in a Two-Electron Double Quantum Dot \\under a Phonon Temperature Gradient}\author{Kazuyuki Kuroyama}
\email[]{kuroyama@iis.u-tokyo.ac.jp}
\affiliation{Institute of Industrial Science, The University of Tokyo,
7-3-1 Komaba, Meguro-ku, Tokyo, 153-8505, Japan}
\affiliation{Center for Emergent Materials Science (CEMS), RIKEN, 2-1 Hirosawa,
Wako-shi, Saitama, 351-0198, Japan}
\author{Sadashige Matsuo}
\affiliation{Center for Emergent Materials Science (CEMS), RIKEN, 2-1 Hirosawa,
Wako-shi, Saitama, 351-0198, Japan}
\author{Seigo Tarucha}
\affiliation{Center for Emergent Materials Science (CEMS), RIKEN, 2-1 Hirosawa,
Wako-shi, Saitama, 351-0198, Japan}
\affiliation{RIKEN Center for Quantum Computing (RQC), RIKEN, 2-1 Hirosawa, 
Wako-shi, Saitama, 351-0198, Japan}
\author{Yasuhiro Tokura}
\email[]{tokura.yasuhiro.ft@u.tsukuba.ac.jp}
\affiliation{Pure and Applied Sciences, University of Tsukuba, 1-1-1 Tennodai, Tsukuba, Ibaraki, 305-8571, Japan}

\begin{abstract}
We have theoretically studied phonon-mediated spin-flip processes of electrons in a GaAs double quantum dot (DQD) holding two spins, under a phonon temperature gradient over the DQD. Transition rates of inter-dot phonon-assisted tunnel processes and intra-dot spin-flip processes involving spin triplet states are formalized by the electron-phonon interaction accompanied with the spin-orbit interaction. The calculations of the spin-flip rates and the occupation probabilities of the spin-states in the two-electron DQD with respect to the phonon temperature difference between the dots are quantitatively consistent with our previous experiment. This theoretical study on the temperature gradient effect onto spins in coupled QDs would be essential for understanding spin-related thermodynamic physics. 
\end{abstract}
\maketitle

\section{Introduction}
Thermodynamics in quantum mesoscopic systems has long been studied as one of the central concepts in solid-state physics. In particular, temperature gradients over mesoscopic structures have realized various kinds of thermodynamic functions such as heat engines and heat valves, which are useful for improvements of thermoelectric conversion and heat control by electrical means in mesoscopic devices \cite{ScheibnerPRL2005, JosefssonNatNano2018, JalielPRL2019, DuttaPRL2020}, and there are many related theoretical studies that have been reported \cite{SanchezPRB2011, JordanPRB2013, StrasbergPRB2018, SanchezPRR2019, ZhangPRE2021, KamimuraPRL2022, YamamotoPRB2022}. Although these kinds of thermodynamic phenomena are constituted mainly by a charge degree of freedom, when spin-related effects such as the spin-orbit interaction exist, contributions of the spin effects to thermodynamic phenomena cannot be dismissed. We have recently realized a lattice temperature gradient over a semiconductor double quantum dot (DQD) by using a nearby QD-based phonon source \cite{KuroyamaPRL2022}. In this experimental study, we revealed that such a phonon source not only generates phonon-mediated spin transitions in the DQD, but also, due to the non-equilibrium phonon environment in the DQD, creates a significant imbalance in the occupation probabilities of parallel and anti-parallel spin configurations in the two-electron DQD. This finding can provide intriguing concepts of spin-dependent thermoelectric devices that are driven by a local lattice temperature gradient. To date many experimental and theoretical studies have been conducted on the phonon-mediated spin-dependent transitions in quantum dots to explore the dephasing mechanism of QD-based spin qubits  \cite{FujisawaNature2002, HansonPRL2005, MeunierPRL2007, GolovachPRL2004, DanonPRB2013}, however, contributions of a temperature gradient over coupled QDs to the transitions have never been discussed substantially. Therefore, theoretical studies on the temperature gradient effect would largely contribute to develop spin-related thermodynamic physics \cite{ScheibnerPRL2005, Hartman2018NatPhys, OnoPRL2020}.

Based on this background, in this work, we have theoretically investigated the phonon-mediated spin-flip processes of two-electron spin states in a GaAs DQD under a phonon temperature gradient to reproduce our previous experiment \cite{KuroyamaPRL2022}. Because spin-orbit interaction and electron-phonon interaction coexist in GaAs, a spin can flip during a transition between ground states and excited states. Furthermore, we considered that the lattice temperatures are different between the two dots, i.e., there is a temperature gradient over the DQD. We successfully formalized the phonon-mediated spin-flip rates in such a non-equilibrium phonon environment. The ratio of the spin-flip tunnel rate from the right to left QD to that in the opposite direction shows clear dependence on the phonon temperature gradient between the QDs. Furthermore, we calculated the occupation probabilities of the two-electron spin states. The significant imbalance between antiparallel and parallel spin states found in our previous experiment is accurately reproduced by the theoretical calculation, which takes into account the non-equilibrium phonon environment in the DQD \cite{KuroyamaPRL2022}.

\section{Hamiltonian of the two-electron DQD system with the spin-orbit interaction}
\begin{figure}[t]
\centering
\includegraphics[width=\linewidth]{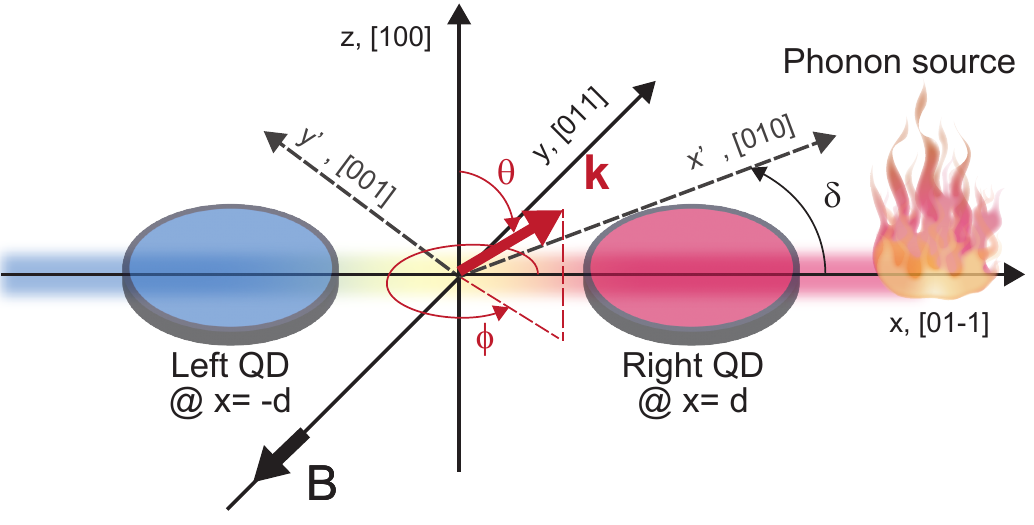}
\caption{ Geometrical configuration of the DQD and the crystallographic orientations. The DQD array is aligned along the [01-1] crystallographic axis, which is denoted by $x$, and each dot is located at $x = \pm d$. The $B$-field is applied in the [0-1-1] direction. The $x'y'$-plane, that is spanned by [010] and [001] crystallographic axis, is used to express the spin-orbit interaction.}
\label{fig1} 
\end{figure}
We considered the two-electron spin configurations of the $(1,1)$ and $(0,2)$ charge states in the DQD, where $(i,j)$ denotes that the electron number in the left and right QDs are $i$ and $j$, respectively. The DQD is formed in a two-dimensional electron gas defined in a GaAs quantum well \cite{HansonRMP2007}. The geometry of the DQD is depicted in Fig. \ref{fig1}. For the parameters defining the DQD, we refer to our previous experiment  \cite{KuroyamaPRL2022}. The two QDs are aligned along the GaAs crystallographic orientation of [01-1]. An in-plane magnetic field $B$ is applied along the [0-1-1] direction. The total Hamiltonian of the two-electron system is
\begin{align}
\hat{\mathcal{H}}=\hat{\mathcal{H}}_{\mathrm{QD\_2e}}+\hat{\mathcal{H}}_{\mathrm{ph}}+\hat{\mathcal{H}}_{\textrm{e-p}},
\end{align}
where $\hat{\mathcal{H}}_{\mathrm{ph}}$ and $\hat{\mathcal{H}}_{\textrm{e-p}}$ are the phonon Hamiltonian and the electron-phonon interaction, respectively. The Hamiltonian of an electron in the DQD system, $\hat{\mathcal{H}}_{\mathrm{QD\_1e}}$, is
\begin{align}
\hat{\mathcal{H}}_{\mathrm{QD\_1e}}&=\hat{\mathcal{H}}_{0}+\hat{\mathcal{H}}_{\mathrm{Z}}+\hat{\mathcal{H}}_{\mathrm{so}}+\hat{\mathcal{H}}_{\mathrm{det}},
\end{align}
where $\hat{\mathcal{H}}_0$ represents the kinetic energy and the potential energy of an electron in the DQD. The Zeeman term and the spin-orbit interaction are denoted as $\hat{\mathcal{H}}_{\mathrm{Z}}$ and $\hat{\mathcal{H}}_{\mathrm{so}}$, respectively. ${\hat{\mathcal{H}}}_{\det} = eEx$ is an energy detuning between the QDs induced by a linear electric field $E$ applied across the DQD. Here, we define $\hat{\mathcal{H}}_0$ in the $xy$-plane as
\begin{align}
\hat{\mathcal{H}}_{0}=-\frac{\hbar^2}{2m^*}\left(\frac{\partial^2}{\partial x^2}+\frac{\partial^2}{\partial y^2}\right)+V(x,y). \nonumber
\end{align}
We consider a parabolic electrostatic potential $V\left( \bm{\mathrm{r}} \right)$, with the local minima at $\bm{\mathrm{r}} = ( \pm d,0)$:
\begin{align}
V(x,y)=\frac{\hbar^2}{2m^*l_0^4}\mathrm{min}\{ (\bm{\mathrm{r}}-\bm{\mathrm{d}})^2,(\bm{\mathrm{r}}+\bm{\mathrm{d}})^2\},
\end{align}
where $l_{0}$ is the confinement length of an electron in the QD, typically 30 to 100 nm, and $m^{*} = 0.067m_{0}$ is the effective mass of electrons in GaAs. Note that the spatial shift of the potential minima induced by ${\hat{\mathcal{H}}}_{\det}$ is dismissed in our calculation \cite{WhitePRB2018}. It also should be noted that we set the external magnetic field $B$ to be sufficiently small ($\sim 0.1$ T) such that a magnetic confinement energy is negligible compared to the original QD confinement energy $\hbar\omega_{0} = \hbar^{2}/(m^{*}l_{0}^{2})$. Therefore, $B$ only defines the spin eigenstates by the Zeeman effect.

The Hamiltonian of the Zeeman term is described as ${\hat{\mathcal{H}}}_{\mathrm{Z}} = \left( g\mu_{B}/2 \right)\bm{\mathrm{B}} \cdot \hat{\bm{\mathrm{\sigma}}}$, where $g$ and $\mu_{B}$ are the Lande g-factor of an electron spin in the QD and the Bohr magneton, respectively. $\hat{\bm{\mathrm{\sigma}}}$ is the vector of the Pauli-spin operators. Furthermore, we make a comment on a local Zeeman energy induced by the Overhauzer field of nuclear spins, inducing the energy difference between $\ket{\uparrow\downarrow(1,1)}$ and $\ket{\downarrow\uparrow(1,1)}$. In previous studies \cite{KoppensScience2005}, a nuclear field is typically 1-5 mT. In such a case, for a weakly coupled DQD, the Zeeman energy difference between the two dots becomes comparable with the exchange energy difference between $|S(1,1)\rangle$ and $|T_{0}(1,1)\rangle$. Therefore, the eigenstates of the anti-parallel spin configurations of the $(1,1)$ charge state are not $\ket{S(1,1)}$ and $\ket{T_0(1,1)}$ but $\ket{\uparrow\downarrow(1,1)}$ and $\ket{\downarrow\uparrow(1,1)}$. In addition, it is assumed that the Zeeman energy originated from the external $B$-field is much stronger than that of the local nuclear field and the exchange energy for $\ket{S(1,1)}$ and $\ket{T_0(1,1)}$. Therefore, it is safely assumed that $\ket{T_{\pm}(1,1)}$ are good quantum basis in this configuration of the weakly coupled DQD system.

The two-electron wavefunctions involved in the phonon-mediated spin-flip processes are defined, by using a single electron wavefunction of the 1s orbital in the left QD, $\Psi_{L}\left( \bm{\mathrm{r}} \right)$, and that of the 1s and 2p orbitals in the right QD, $\Psi_{R}\left( \bm{\mathrm{r}} \right)$, and $\Psi_{R^{*}}\left( \bm{\mathrm{r}} \right)$ (see APPENDIX A for the detailed expressions of the wavefunctions):
\begin{align}
\Psi_{\uparrow\downarrow(1,1)}\equiv&\frac{1}{\sqrt{2}}\left(\Psi_L(\bm{\mathrm{r}}_1)\Psi_R(\bm{\mathrm{r}}_2)\xi_{\uparrow 1}\xi_{\downarrow 2}\right.\nonumber\\
&\left.-\Psi_R(\bm{\mathrm{r}}_1)\Psi_L(\bm{\mathrm{r}}_2)\right)\xi_{\downarrow 1}\xi_{\uparrow 2},\\
\Psi_{\downarrow\uparrow(1,1)}\equiv&\frac{1}{\sqrt{2}}\left(\Psi_R(\bm{\mathrm{r}}_1)\Psi_L(\bm{\mathrm{r}}_2)\xi_{\uparrow 1}\xi_{\downarrow 2}\right.\nonumber\\
&\left.-\Psi_L(\bm{\mathrm{r}}_1)\Psi_R(\bm{\mathrm{r}}_2)\right)\xi_{\downarrow 1}\xi_{\uparrow 2},\\
\Psi_{T\pm(1,1)}\equiv&\frac{1}{\sqrt{2}}\left(\Psi_L(\bm{\mathrm{r}}_1)\Psi_R(\bm{\mathrm{r}}_2)-
\Psi_R(\bm{\mathrm{r}}_1)\Psi_L(\bm{\mathrm{r}}_2)\right)\xi_{T\pm},\\
\Psi_{S(0,2)}\equiv&\Psi_R(\bm{\mathrm{r}}_1)\Psi_R(\bm{\mathrm{r}}_2)\xi_S,\\
\Psi_{T\nu(0,2)}\equiv&\frac{1}{\sqrt{2}}\left(\Psi_R(\bm{\mathrm{r}}_1)\Psi_{R^*}(\bm{\mathrm{r}}_2)
-\Psi_{R^*}(\bm{\mathrm{r}}_1)\Psi_R(\bm{\mathrm{r}}_2)\right)\xi_{T\nu},
\end{align}
where $T$ with $\nu = \pm ,0$ and $S$ denote the spin triplet and singlet states, respectively, and $\xi_{S}$ and $\xi_{T\nu}$ represent the corresponding spin wavefunctions. To describe the spin wavefunctions, we define $\xi_{\uparrow i}$ and $\xi_{\downarrow i}$ for $i = 1,2$, which denote the spin wavefunctions of the $i$-th electron spin and satisfy $\xi_{\uparrow i}^{\dagger}\xi_{\uparrow j} = \xi_{\downarrow i}^{\dagger}\xi_{\downarrow j} = \delta_{ij}$ and $\xi_{\uparrow i}^{\dagger}\xi_{\downarrow j} = 0$. Then, the spin wavefunctions of the spin singlet $\xi_{S}$ and triplet states $\xi_{T\nu}$ are described as follows.
\begin{align}
\xi_S&\equiv\frac{1}{\sqrt{2}}\left(\xi_{\uparrow 1}\xi_{\downarrow 2}
-\xi_{\downarrow 1}\xi_{\uparrow 2}\right),\\
\xi_{T+}&=\xi_{\uparrow 1}\xi_{\uparrow 2},\\
\xi_{T0}&=\frac{1}{\sqrt{2}}\left(\xi_{\uparrow 1}\xi_{\downarrow 2}
+\xi_{\downarrow 1}\xi_{\uparrow 2}\right),\\
\xi_{T-}&=\xi_{\downarrow 1}\xi_{\downarrow 2}.
\end{align}
Note that the spin-quantization axis is selected to be parallel to the in-plane magnetic field, $\bm{\mathrm{B}}$, (see Fig. \ref{fig1}(a)). The two-electron Hamiltonian without the spin-orbit interaction is $\hat{\mathcal{H}}_{0\mathrm{Z\_2e}}\equiv \sum_{i=1,2} \left\{\hat{\mathcal{H}}_0^{(i)} +\hat{\mathcal{H}}_{\mathrm{Z}}^{(i)}+\hat{\mathcal{H}}_{\mathrm{det}}^{(i)}\right\}+\hat{\mathcal{H}}_{\mathrm{int}}$, where $\hat{\mathcal{H}}_{\mathrm{int}}$ represents the Coulomb interaction. Then, the diagonal matrix elements of $\hat{\mathcal{H}}_{0\mathrm{Z\_2e}}$ are
\begin{align}
\left\langle \uparrow\downarrow(1,1)\left|\hat{\mathcal{H}}_{\mathrm{0Z\_2e}}\right|\uparrow\downarrow(1,1)\right\rangle
=&\epsilon_L+\epsilon_R+V,\\
\left\langle \downarrow\uparrow(1,1)\left|\hat{\mathcal{H}}_{\mathrm{0Z\_2e}}\right|\downarrow\uparrow(1,1)\right\rangle
=&\epsilon_L+\epsilon_R+V,\\
\left\langle T_{\pm}(1,1)\left|\hat{\mathcal{H}}_{\mathrm{0Z\_2e}}\right|T_{\pm}(1,1)\right\rangle
=&\epsilon_L+\epsilon_R+V\pm g\mu_{\mathrm{B}}B,\\
\left\langle S(0,2)\left|\hat{\mathcal{H}}_{\mathrm{0Z\_2e}}\right|S(0,2)\right\rangle=&2\epsilon_R+\epsilon_{\mathrm{det}}+U,\\
\left\langle T_0(0,2)\left|\hat{\mathcal{H}}_{\mathrm{0Z\_2e}}\right|T_0(0,2)\right\rangle
=&\epsilon_R+\epsilon_{R^*}+\epsilon_{\mathrm{det}}+U',\\
\left\langle T_{\pm}(0,2)\left|\hat{\mathcal{H}}_{\mathrm{0Z\_2e}}\right|T_{\pm}(0,2)\right\rangle
=&\epsilon_R+\epsilon_{R^*}+\epsilon_{\mathrm{det}}\nonumber\\
&+U'\pm g\mu_{\mathrm{B}}B.
\end{align}
where $V$ and $U$ are the inter- and intra-QD charging energy, respectively. $U' = U - K$, where $K\ ( > 0)$ is the change in the Coulomb interaction energy between $\ket{S(0,2)}$ and $\ket{T_{\nu}(0,2)}$ \cite{HansonRMP2007}. $\epsilon_{L}$ and $\epsilon_{R}$ are the eigenenergies of the left and right QDs, and $\epsilon_{R^{*}}$ is that of the first excited state in the right QD (see APPENDIX A). $\epsilon_{\det} \simeq 2eEd$ is the energy detuning between the $(1,1)$ and $(0,2)$ charge states. Note that the dependence of the intra-QD charging energy on the wavefunctions $\Psi_{L}$, and $\Psi_{R}$ is neglected. The off-diagonal matrix elements of ${\hat{\mathcal{H}}}_{\mathrm{0Z\_2e}}$, which represent the inter-dot tunnel coupling, are
\begin{align}
\left\langle \uparrow\downarrow(1,1)\left|\hat{\mathcal{H}}_{\mathrm{0Z\_2e}}\right|S(0,2)\right\rangle=&\left\langle \downarrow\uparrow(1,1)\left|\hat{\mathcal{H}}_{\mathrm{0Z\_2e}}\right|S(0,2)\right\rangle\nonumber\\
=&\tau_{LR},\\
\left\langle T_\nu(1,1)\left|\hat{\mathcal{H}}_{\mathrm{0Z\_2e}}\right|T_\nu(0,2)\right\rangle=&-\tau_{LR^*},
\end{align}
for $\nu=\pm$, and 
\begin{align}
    \left\langle \uparrow\downarrow(1,1)\left|\hat{\mathcal{H}}_{\mathrm{0Z\_2e}}\right|T_0(0,2)\right\rangle=&-\left\langle \downarrow\uparrow(1,1)\left|\hat{\mathcal{H}}_{\mathrm{0Z\_2e}}\right|T_0(0,2)\right\rangle\nonumber\\
    =&-\tau_{LR^*}/\sqrt{2}.
\end{align}
The definitions of $\tau_{LR}$ and $\tau_{LR^{*}}$ are described in APPENDIX A. Finally, the Hamiltonian of the spin-orbit interaction is
\begin{align}
\hat{\mathcal{H}}_{\mathrm{so}}=&\frac{\hbar}{m^*\lambda_{\mathrm{D}}}
(-\hat{p}_{x'}\hat{\sigma}_{x'}+\hat{p}_{y'}\hat{\sigma}_{y'})\nonumber\\
&+\frac{\hbar}{m^*\lambda_{\mathrm{R}}}(\hat{p}_{x'}\hat{\sigma}_{y'}-\hat{p}_{y'}\hat{\sigma}_{x'}),
\end{align}
where $\lambda_{\mathrm{D}}$ and $\lambda_{\mathrm{R}}$ are the spin-orbit lengths for the Dresselhaus and Rashba interactions, respectively. 
The $x'$ and $y'$ axis are set along [010] and [001] in GaAs, respectively, as depicted in Fig. \ref{fig1}. The angle between the $x$ and $x'$ axis is denoted by $\delta$. The coordinates of the $xy$ and $x'y'$-planes can be transformed as $(x',y') = \left( x\cos\delta + y\sin\delta, - x\sin\delta + y\cos\delta \right)$, using the angle $\delta$. In the $xy$-plane, the spin-orbit interaction Hamiltonian is simply given by (see APPENDIX B for the derivation)
\begin{align}
\hat{\mathcal{H}}_{\mathrm{so}}&=-\frac{\hbar}{m^*}\left[\frac{\cos2\delta}{\lambda_{\mathrm{D}}}\hat{p}_{x}\hat{\sigma}_{y}+\left(\frac{\sin2\delta}{\lambda_{\mathrm{D}}}-\frac{1}{\lambda_{\mathrm{R}}}\right)\hat{p}_{y}\hat{\sigma}_{y}\right].\label{eq:SOI}
\end{align}
In our previous experimental study in \cite{KuroyamaPRL2022}, $\delta = \pi/4$ was chosen. In this condition, the off-diagonal elements involved in the spin-orbit interaction can be described by
\begin{align}
\left\langle T_{\pm}(1,1)\left| \sum_{i = 1,2}^{}{\hat{\mathcal{H}}}_{\mathrm{so}}^{(i)} \right|S(0,2) \right\rangle & = \frac{\hbar^{2}}{m^{*}}\frac{d_{\mathrm{ID}}}{\lambda_{\mathrm{so}}}. 
\end{align}
where $i=$ 1,2, which represents the individual electrons in the DQD. The effective spin-orbit inverse-length is defined as $\lambda_{\mathrm{so}}^{-1}\equiv \lambda_{\mathrm{R}}^{-1} - \lambda_{\mathrm{D}}^{-1}$. The inter-dot inverse dipole element, $d_{\mathrm{ID}}$, is given as
\begin{align}
d_{\mathrm{ID}}\equiv \int d^2\bm{\mathrm{r}}\Psi_L^*(\bm{\mathrm{r}})\frac{\partial}{\partial y}\Psi_R(\bm{\mathrm{r}}).
\end{align}
Similarly, the off-diagonal elements of $\hat{\mathcal{H}}_{\mathrm{so}}$ between $\ket{T_{\pm}(0,2)}$ and $\ket{S(0,2)}$ are
\begin{align}
\left\langle T_{\pm}(0,2)\left| \sum_{i = 1,2}^{}{\hat{\mathcal{H}}}_{\mathrm{so}}^{(i)} \right|S(0,2) \right\rangle = - \frac{\hbar^{2}}{m^{*}}\frac{d_{R}}{\lambda_{\mathrm{so}}},
\end{align}
where the inverse dipole matrix element in the right QD, $d_R$, is
\begin{align}
d_{R}\equiv\int d^2\bm{\mathrm{r}}\Psi_{R^*}^*(\bm{\mathrm{r}})\frac{\partial}{\partial y}\Psi_R(\bm{\mathrm{r}}).
\end{align}
To summarize the above discussion, the Hamiltonian matrix of the two electrons in the laterally coupled DQD are expressed in terms of the basis states, $(\ket{\uparrow\downarrow(1,1)}$; $\ket{\downarrow\uparrow(1,1)}$; $\ket{T_+(1,1)}$; $\ket{T_-(1,1)}$; $\ket{S(0,2)}$; $\ket{T_+(0,2)}$; $\ket{T_-(0,2)}$; $\ket{T_0(0,2)}$):
\begin{align}
\hat{\mathcal{H}}_{\mathrm{QD\_2e}}=&\sum_{i = 1,2}^{}\left\{ {\hat{\mathcal{H}}}_{0}^{(i)} + {\hat{\mathcal{H}}}_{\mathrm{Z}}^{(i)} + {\hat{\mathcal{H}}}_{\mathrm{so}}^{(i)} + {\hat{\mathcal{H}}}_{\det}^{(i)} \right\} + {\hat{\mathcal{H}}}_{int}\nonumber\\
=&\left(
  \begin{matrix}
    0 &0 & 0 & 0   \\
    0 &0 & 0 & 0   \\
    0 &0 & E_{\mathrm{\mathrm{Z}}} & 0 \\
    0 &0 & 0 & -E_{\mathrm{Z}}\\
    \tau_{LR}^* &\tau_{LR}^* & \alpha^* & \alpha^*  \\
    0 &0 & -\tau_{LR^*}^* & 0  \\
    0 &0 & 0 & -\tau_{LR^*}^*  \\
    -\tau_{LR^*}^*/\sqrt{2} & \tau_{LR^*}^*/\sqrt{2} & 0 & 0     
  \end{matrix}\right.\nonumber
  \\
&\left.
  \begin{matrix}
    \tau_{\mathrm{LR}} & 0 & 0 & -\tau_{LR^*}/\sqrt{2}\\
    \tau_{\mathrm{LR}} & 0 & 0 & \tau_{LR^*}/\sqrt{2}\\
    \alpha & -\tau_{LR^*} & 0 & 0\\
    \alpha & 0 & -\tau_{LR^*} & 0\\
    \varepsilon & \beta & \beta & 0\\
    \beta^* & \epsilon'+E_{\mathrm{Z}} & 0 & 0\\
    \beta^* & 0 & \epsilon'-E_{\mathrm{Z}} &0\\\
    0 & 0 & 0 & \epsilon'
  \end{matrix}\right),\label{eq:Hamiltonian}
\end{align}
where the diagonal elements are expressed as the difference from the reference energy, $\epsilon_{L} + \epsilon_{R} + V$. The Zeeman energy is described as $E_{\mathrm{Z}} \equiv g\mu_{B}B$. $\epsilon \equiv \epsilon_{R} - \epsilon_{L} + \epsilon_{\det} + U - V$ is the energy detuning between $\ket{\uparrow\downarrow(1,1)}$ and $\ket{S(0,2)}$ and between $\ket{\downarrow\uparrow(1,1)}$ and $\ket{S(0,2)}$, respectively and $\epsilon' \equiv \epsilon_{R^{*}} - \epsilon_{L} + \epsilon_{\det} + U' - V$ is that between $\ket{\uparrow\downarrow(1,1)}$and $\ket{T_0(0,2)}$ and between $\ket{\downarrow\uparrow(1,1)}$ and $\ket{T_0(0,2)}$ Note that there is a local Zeeman energy difference between $\ket{\uparrow \downarrow (1,1)}$ and $\ket{\downarrow \uparrow (1,1)}$, but since it is assumed that the local Zeeman energy difference is much smaller than $E_{\mathrm{Z}}$ and $\Delta$, it can be negligible in the Hamiltonian ${\hat{\mathcal{H}}}_{\mathrm{QD\_ 2e}}$ and in the following calculation. The off-diagonal elements are $\alpha \equiv \hbar^{2}d_{\mathrm{ID}}/\left( m^{*}\lambda_{\mathrm{so}} \right)$ and $\beta \equiv - \hbar^{2}d_{R}/\left( m^{*}\lambda_{\mathrm{so}} \right)$. The $\ket{T_{0}(0,2)}$ is only accessed from $\ket{\uparrow \downarrow (1,1)}$ and $\ket{\downarrow \uparrow (1,1)}$ by the phonon-assisted tunneling process. Note that in the following derivations, we consider only the resonant condition where the energies of $\ket{S(0,2)}$ and $\ket{\uparrow \downarrow (1,1)}$ are equivalent, i.e., $\epsilon = 0$, by choosing $\epsilon_{\det}$ appropriately (see Fig. \ref{fig2}(a)). Furthermore, in such a case, the energy separation between $\ket{S(0,2)}$ and $\ket{T_{0}(0,2)}$ is described as $\Delta = \epsilon' = \epsilon_{R^{*}} - \epsilon_{R} - K$.

By diagonalizing the matrix in Eq. \eqref{eq:Hamiltonian}, we obtain the eigenenergies and corresponding two-electron wavefunctions: ${\hat{\mathcal{H}}}_{\mathrm{QD\_ 2e}}\ket{\xi_{n}} = E_{n} \ket{\xi_{n}}$, where the eigenfunctions can be expanded as $\ket{\xi_{n}} = \sum_{j = 1}^{8}C_{nj} \ket{\psi_{j}}$. Since the inter-QD tunnel coupling and spin-orbit interactions are relatively weak, the off-diagonal elements of ${\hat{\mathcal{H}}}_{\mathrm{QD\_ 2e}}$ are assumed to be very small. Therefore, the first-order perturbation theory can be applied to obtain the two-electron wavefunctions: $C_{n,j} \simeq \langle\psi_{j}| {\hat{\mathcal{H}}}_{1}|\psi_{n}\rangle/\left( E_{n}^{0} - E_{j}^{0} \right)$ for $n \neq j$, and $C_{n,n}=1$ where ${\hat{\mathcal{H}}}_{1}$ is a matrix comprising the off-diagonal elements of ${\hat{\mathcal{H}}}_{\mathrm{QD\_ 2e}}$, and $E_n^0$ is the $n$-th diagonal element of ${\hat{\mathcal{H}}}_{\mathrm{QD\_ 2e}}$. Hence please note that $|\xi_{1}\rangle$, $\ket{\xi_{2}} \simeq$ $\ket{\uparrow \downarrow (1,1)}$, $\ket{\downarrow \uparrow (1,1)}$, and $\ket{\xi_{3}}$, $\ket{\xi_{4}}$ $\simeq$ $\ket{T_{+}(1,1)}$, $\ket{T_{-}(1,1)}$. $\ket{\xi_{5}}$ $\simeq \ket{S(0,2)}$, and $\ket{\xi_{6}}$, $\ket{\xi_{7}}$, $\ket{\xi_{8}}$ $\simeq \ket{T_{+}(0,2)}$, $\ket{T_{-}(0,2)}$, $\ket{T_{0}(0,2)}$.

\section{Electron-phonon interaction}
\begin{figure}[t]
\centering
\includegraphics[width=\linewidth]{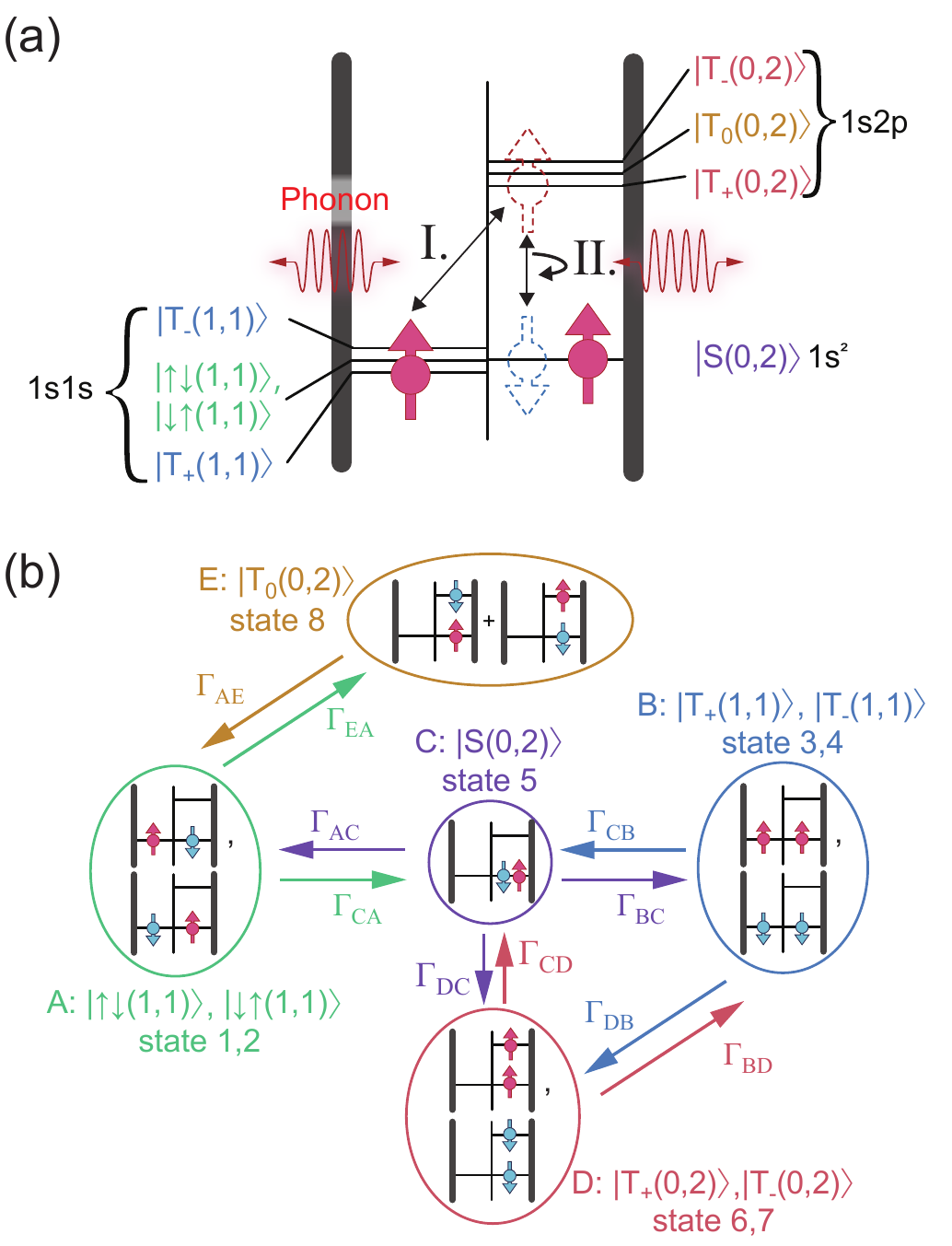}
\caption{(a) Energy diagram which explains the phonon-induced spin-flip tunnel processes in the two-electron DQD. A phonon excitation from $\ket{T_{+}(1,1)}$ to the excited state of $\ket{T_{+}(0,2)}$ (Process I), which is followed by the intra-dot spin-flip relaxation processes from $\ket{T_{+}(0,2)}$ to $\ket{S(0,2)}$ (Process II), is illustrated. (b) Transition diagram of the two-electron spin states with phonon excitations.
}
\label{fig2} 
\end{figure}
Next, we introduce the electron-phonon interaction. Figure \ref{fig2}(a) shows the energy diagram of the two-electron spin states in a DQD, which are involved in the phonon-mediated spin-flip tunnel processes. For the initial state, $\ket{T_{+}(1,1)}$, the electron residing in the left QD can be excited to the 2p orbital in the right QD, and the two electrons form $\ket{T_{+}(0,2)}$. Secondly, after this phonon-assisted inter-dot tunnel, the electron in the 2p orbital rapidly relaxes to the 1s orbital in the right QD to form $\ket{S(0,2)}$, accompanied by the spin flip and the phonon emission.

To estimate these phonon-mediated transition rates, we take into account the electron-phonon interaction. The Hamiltonian of the electron-phonon interaction is
\begin{align}
\hat{\mathcal{H}}_{\textrm{e-p}}&=\sum_{\mu,\bm{\mathrm{k}}}\sum_{ij}\Lambda_{\mu,\bm{\mathrm{k}},ij}
\ket{\xi_i}\bra{\xi_j}\left(\hat{b}_{\mu,\bm{\mathrm{k}}}^\dagger+\hat{b}_{\mu,-\bm{\mathrm{k}}}\right),
\end{align}
where $\mu$ and $\bm{\mathrm{k}}$ represent the mode and the wave number of a phonon, respectively, and $\hat{b}_{\mu,\bm{\mathrm{k}}}$ and $\hat{b}_{\mu,\bm{\mathrm{k}}}^\dagger$ are the corresponding annihilation and creation operators satisfying the commutation relation: $\left[\hat{b}_{\mu,\bm{\mathrm{k}}},\hat{b}_{\mu',\bm{\mathrm{k}}'}^\dagger\right]=\delta_{\mu,\mu'}\delta_{\bm{\mathrm{k}},\bm{\mathrm{k}}'}$. $\Lambda_{\mu,\bm{\mathrm{k}},ij}\equiv \lambda_{\mu,\bm{\mathrm{k}}} \braket{\xi_i|e^{i\bm{\mathrm{k}}\cdot\bm{\mathrm{r}}_1}+e^{i\bm{\mathrm{k}}\cdot\bm{\mathrm{r}}_2}|\xi_j}$. $\lambda_{\mu,\bm{\mathrm{k}}}$ is the electron-phonon coupling parameter. Based on the Fermi's golden rule (in other words, second-order perturbation theory or Lindbald-type master equation formalism), the transition rate from level $n$ to level $m$ via absorption or emission of one phonon is
\begin{align}
\gamma_{mn}&\equiv \frac{2\pi}{\hbar}\sum_{\mu,\bm{\mathrm{k}}}\left|\Lambda_{\mu,\bm{\mathrm{k}},mn}\right|^2\nonumber\\
&\times\left[ n_{\mu,\bm{\mathrm{k}}}^{a}\delta(\epsilon_n-\epsilon_m+\hbar\omega_{\mu,\bm{\mathrm{k}}})\right.\nonumber\\
&\left.+(n_{\mu,\bm{\mathrm{k}}}^{a}+1)\delta(\epsilon_n-\epsilon_m-\hbar\omega_{\mu,\bm{\mathrm{k}}})\right],
\end{align}
where $n_{\mu,k}^{a}$ is the phonon distribution function contributing to the processes $a=\ket{\xi_3}\leftrightarrow\ket{\xi_6}$ and $\ket{\xi_4}\leftrightarrow\ket{\xi_7}$ ($\sim \ket{T_{\pm}(1,1)} \leftrightarrow \ket{T_{\pm}(0,2)}$) and $a=\ket{\xi_5}\leftrightarrow\ket{\xi_6}$ and $\ket{\xi_5}\leftrightarrow\ket{\xi_7}$ ($\sim \ket{S(0,2)} \leftrightarrow \ket{T_{\pm}(0,2)}$ ). In the global equilibrium condition, based on spatial uniformity, $n_{\mu,k}(T)\equiv 1/(e^{\hbar\omega_{\mu,\bm{\mathrm{k}}}/(k_{\mathrm{B}}T)}-1)$. Here, we introduce individual phonon temperatures $T_L$ for the transition $\ket{\xi_3}\leftrightarrow\ket{\xi_6}$ (and $\ket{\xi_4}\leftrightarrow\ket{\xi_7}$), and  $T_R$ for the transition $\ket{\xi_5}\leftrightarrow\ket{\xi_6}$ (and the transition $\ket{\xi_5}\leftrightarrow\ket{\xi_7}$). The transition rates between states $\ket{\xi_5}$ and $\ket{\xi_6}$, and those between states $\ket{\xi_3}$ and $\ket{\xi_6}$ are described by using the phonon temperatures:
\begin{align}
\Gamma_{56}=&\gamma_0(1+n_{\mu,k_{\Delta}}(T_R)),\label{eq:g56}\\
\Gamma_{65}=&\gamma_0n_{\mu,k_{\Delta}}(T_R),\\ 
\Gamma_{36}=&\gamma_0'(1+n_{\mu,k_{\Delta}}(T_L)),\label{eq:g36}\\
\Gamma_{63}=&\gamma_0'n_{\mu,k_{\Delta}}(T_L),
\end{align}
where $k_{\Delta}\equiv \Delta/(\hbar c_{\mu})$, and $\Gamma_{ij}$ denotes the transition rate from state $j$ to $i$. The coefficients, $\gamma_0$ and $\gamma_0'$, in Eqs. \eqref{eq:g56} and \eqref{eq:g36} are derived by the following calculations.
\begin{align}
\gamma_0&\equiv\frac{2\pi}{\hbar}\sum_{\mu,\bm{\mathrm{k}}}\left|\Lambda_{\mu,\bm{\mathrm{k}},56}\right|^2
\delta(\Delta-\hbar\omega_{\mu,\bm{\mathrm{k}}})\nonumber\\
&\simeq\frac{2\pi}{\hbar}\sum_{\mu}\left\langle\left|\Lambda_{\mu,\bm{\mathrm{k}},56}\right|^2\right\rangle_{k_{\Delta}}
\nu_{\mu}(k_{\Delta}),\label{eq:gamma_0}\\ 
\gamma_0'&\equiv\frac{2\pi}{\hbar}\sum_{\mu,\bm{\mathrm{k}}}\left|\Lambda_{\mu,\bm{\mathrm{k}},36}\right|^2
\delta(\Delta-\hbar\omega_{\mu,\bm{\mathrm{k}}})\nonumber\\
&\simeq \frac{2\pi}{\hbar}\sum_{\mu}\left\langle\left|\Lambda_{\mu,\bm{\mathrm{k}},36}\right|^2\right\rangle_{k_{\Delta}}
\nu_{\mu}(k_{\Delta}),\label{eq:gamma_0_d}
\end{align}
where $\nu_{\mu}(k)\equiv Vk^2/(2\pi^2\hbar c_{\mu})$ and $c_{\mu}$ are the phonon density of states of mode $\mu$ and a sound velocity of $\mu$, respectively. $\mu$ denotes either a longitudinal acoustic phonon ($\mu=l$) or a transverse acoustic phonon ($\mu=t$). $V=L^3$ is the system volume. The square of the phonon matrix element, $\left\langle\left|\Lambda_{\mu,\bm{\mathrm{k}},ij}\right|^2\right\rangle_{k_{\Delta}}$, is approximated by its average value over the surface of the wavenumber space defined as $\left|\bm{\mathrm{k}}\right|=k_{\Delta}$:
\begin{align}
\left\langle\left|\Lambda_{\mu,\bm{\mathrm{k}},ij}\right|^2\right\rangle_{k_{\Delta}}
&\equiv \frac{1}{4\pi}\int_0^{2\pi}d\phi \int_0^{\pi}d\theta\sin\theta\nonumber\\
&\times \left.\left|\Lambda_{\mu,\bm{\mathrm{k}},ij}\right|^2\right|_{\left|\bm{\mathrm{k}}\right|=k_{\Delta}}\label{eq:el-ph_int},
\end{align}
where the orientation of the wavenumber, $\bm{\mathrm{k}}/k=(\sin\theta\cos\phi,\sin\theta\sin\phi,\cos\theta)$, is represented by the spherical coordinate (see a red arrow in Fig. \ref{fig1} for the definition of $\bm{\mathrm{k}}$).
\begin{table*}[t]
\begin{tabular}{|l|l|}
\hline
\hline
Mass density of GaAs, $\rho$&  5300 kg/m\\ 
Speed of sound for longitudinal phonons in GaAs, $c_l$ &  4700 m/sec\\ 
Deformation potential of GaAs, $\Xi$ \cite{DanonPRB2013} &  13.7 eV\\ 
Electron Lande $g$-factor &  -0.44\\
Quantum well length, $l_{z}$ &  15 nm\\ 
Magnetic field, $B$  &  0.1 T\\ 
Spin-orbit length, $\lambda_{\mathrm{so}}$ \cite{MaisiPRL2016, FujitaPRL2016, HoffmannPRL2017}&1 $\mathrm{\mu m}$\\
\hline
\hline
\end{tabular}
\caption{Values of the constants used in the calculations.}\label{Tab1}
\end{table*}

For the derivation of the phonon matrix elements $\Lambda_{\mu,\bm{\mathrm{k}},ij}$, we calculate the bracket product of $\braket{\xi_i|e^{i\bm{\mathrm{k}}\cdot\bm{\mathrm{r}}}|\xi_j}$ based on the wavefunctions perturbed by the spin-orbit interaction. Phonons do not alter the electron spin states and solely interact with electrons individually; therefore, the non-zero matrix elements of $\hat{\mathcal{O}}_{\bm{\mathrm{k}}}\equiv \sum_{i=1,2}e^{i\bm{\mathrm{k}}\cdot\bm{\mathrm{r}}_i}$, i.e., the main part of the electron-phonon interaction Hamiltonian, is
\begin{align}
\left\langle \xi_n\left|\hat{\mathcal{O}}_{\bm{\mathrm{k}}}\right|\xi_m\right\rangle&=
\sum_{j=1,j'=1}^8C_{nj}^*C_{mj'}\braket{\psi_j|\hat{\mathcal{O}}_{\bm{\mathrm{k}}}|\psi_{j'}}.
\end{align}
At the weak coupling regime of a DQD, the off-diagonal components can be neglected because they are much smaller by a factor of $\mathrm{e}^{-d^2/l_0^2}$ than the diagonal components. Therefore, using only the diagonal components, we can express the matrix elements explicitly as
\begin{align}
\left\langle \xi_n\left|\hat{\mathcal{O}}_{\bm{\mathrm{k}}}\right|\xi_m\right\rangle&\simeq	
\sum_{j=1}^8 C_{nj}^*C_{mj}O_R\nonumber\\
&+\sum_{j=1}^4 C_{nj}^*C_{mj}O_L\nonumber\\
&+C_{n5}^*C_{m5}O_R\nonumber\\
&+\sum_{j=6}^8 C_{nj}^*C_{mj}O_{R^*},
\end{align}
where $O_{L}\equiv \braket{\Psi_L|e^{i\bm{\mathrm{k}}\cdot\bm{\mathrm{r}}}|\Psi_L}$, $O_{R}\equiv \braket{\Psi_R|e^{i\bm{\mathrm{k}}\cdot\bm{\mathrm{r}}}|\Psi_R}$ and $O_{R^*}\equiv \braket{\Psi_{R^*}|e^{i\bm{\mathrm{k}}\cdot\bm{\mathrm{r}}}|\Psi_{R^*}}$. Based on the above results, the off-diagonal element between states $\ket{\xi_6}\ (\simeq\ket{T_+(0,2)})$ and $\ket{\xi_5}$ $(\simeq\ket{S(0,2)})$ is
\begin{align}
\braket{\xi_6|\hat{\mathcal{O}}_{\bm{\mathrm{k}}}|\xi_5}
\simeq &\frac{\alpha\tau_{LR^*}}{\Delta E_{\mathrm{Z}}}(O_L+O_{R})+\frac{\beta}{\Delta+E_{\mathrm{Z}}}(O_R-O_{R^*})\nonumber\\
\simeq &\frac{\alpha\tau_{LR^*}}{\Delta E_{\mathrm{Z}}}(O_L+O_{R})+\frac{\beta}{\Delta}(O_R-O_{R^*}), \label{eq:O65}
\end{align}
where it is assumed that $E_{\mathrm{Z}}\ll \Delta$ from our previous experimental condition \cite{KuroyamaPRL2022} of $B=0.1$ T.  For the weak inter-dot coupling condition, since $\alpha$ is much smaller than $E_{\mathrm{Z}}$, the first term in Eq. \eqref{eq:O65} can be ignored as
\begin{align}
    \braket{\xi_6|\hat{\mathcal{O}}_{\bm{\mathrm{k}}}|\xi_5}\simeq\frac{\beta}{\Delta}(O_R-O_{R^*}).
\end{align}
Similarly, the off-diagonal element between states $\ket{\xi_6}$ $(\simeq\ket{T_+(0,2)})$ and $\ket{\xi_3}$ $(\simeq\ket{T_+(1,1)})$ is
\begin{align}
\braket{\xi_3|\hat{\mathcal{O}}_{\bm{\mathrm{k}}}|\xi_6}
\simeq&-\frac{\tau^*_{LR^*}}{\Delta}(O_L-O_{R^*}).
\end{align}
Regarding the $T_-$ states, the off-diagonal elements between states $\ket{\xi_7}\ (\simeq\ket{T_-(0,2)})$ and $\ket{\xi_5}\ (\simeq\ket{S(0,2)})$ and between states $\ket{\xi_4}$ $(\simeq\ket{T_-(1,1)})$ and $\ket{\xi_7}$ $(\simeq\ket{T_-(0,2)})$ can also be determined:
\begin{align}
\braket{\xi_7|\hat{\mathcal{O}}_{\bm{\mathrm{k}}}|\xi_5}&\simeq\frac{\beta}{\Delta}(O_R-O_{R^*}),\\
\braket{\xi_4|\hat{\mathcal{O}}_{\bm{\mathrm{k}}}|\xi_7}&\simeq-\frac{\tau^*_{LR^*}}{\Delta}(O_L-O_{R^*}).
\end{align}
Similarly, regarding the $T_0$ states, the off-diagonal elements between states $\ket{\xi_1}$and $\ket{\xi_8}$ and between states $\ket{\xi_2}$ and $\ket{\xi_8}$ are described as $\braket{\xi_1|\hat{\mathcal{O}}_{\bm{\mathrm{k}}}|\xi_8} \simeq -\braket{\xi_2|\hat{\mathcal{O}}_{\bm{\mathrm{k}}}|\xi_8} \simeq -\tau^*_{LR^*}(O_L-O_{R^*})/(\sqrt{2}\Delta)$.

Next, we introduce the electron-phonon coupling coefficient $\lambda_{\mu,\bm{\mathrm{k}}}$, into the calculation of $\Lambda_{\mu,\bm{\mathrm{k}},ij}$. The possible contributions in GaAs QDs are related to the piezo-electric effect and the deformation potential. We confirmed that, for a phonon energy around $\Delta\sim 1$ meV, the electron-phonon coupling is dominated by the deformation potential \cite{GolovachPRL2004}, and the piezo-electric effect can be neglected \cite{DanonPRB2013}. Therefore, we only considered the deformation potential. The electron-phonon coupling coefficient of the deformation potential is described by $\lambda_{\mu=l,\bm{\mathrm{k}}}=\sqrt{\hbar/(2V\rho c_l})(-i\sqrt{|\bm{\mathrm{k}}|}\Xi)$, where $\rho$ is the mass density of GaAs, and $c_l$ is the velocity of a longitudinal acoustic phonon. $\Xi$ is the deformation potential of GaAs. In the following calculations, only longitudinal acoustic phonons ($\mu=l$) with the energy $\Delta$ is considered.

\section{Evaluation of the phonon-mediated spin-flip tunnel rates}
\begin{figure*}[t]
\centering
\includegraphics[width=\linewidth]{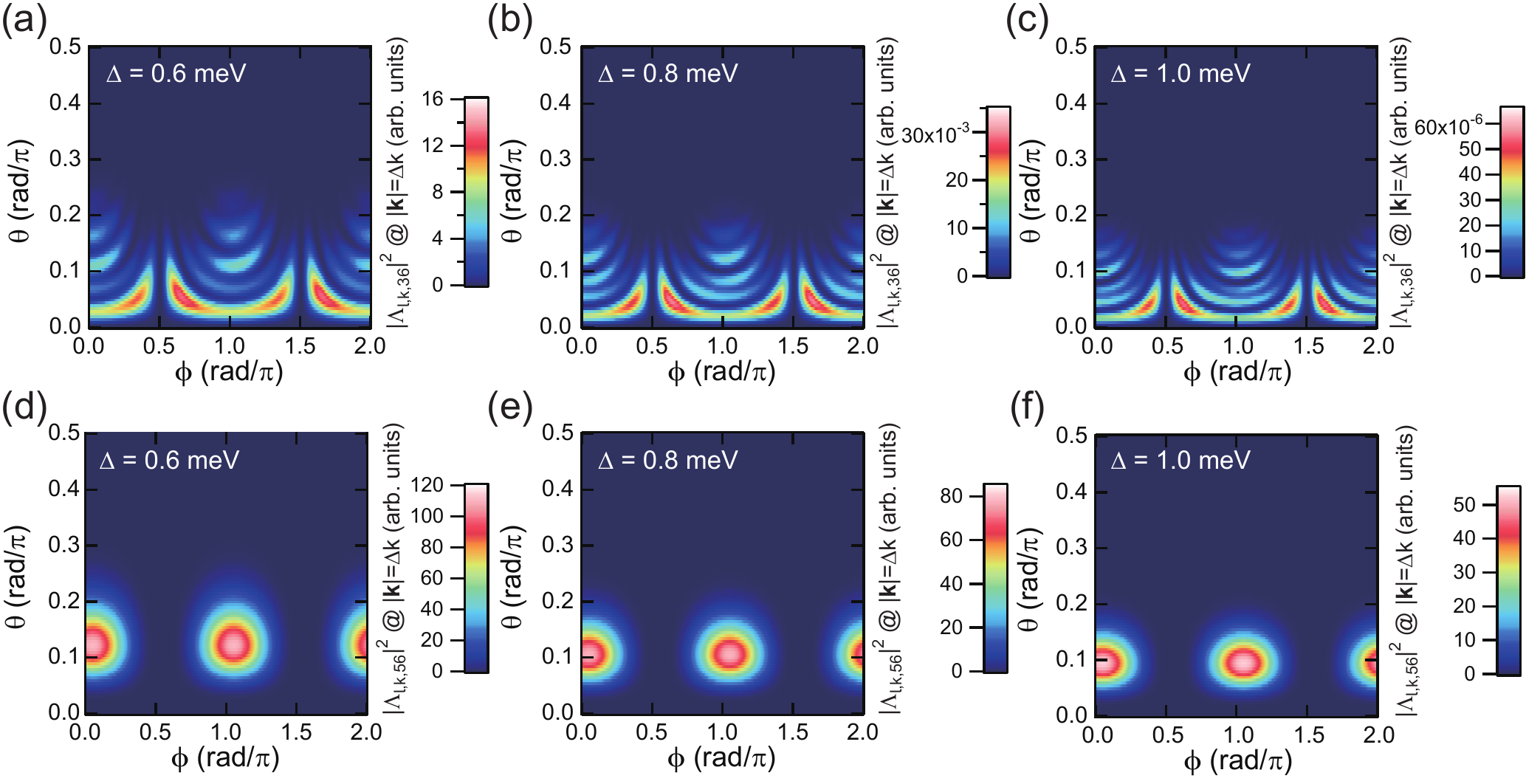} 
\caption{(a)-(c) Color-coded maps of $\left| \Lambda_{l,\bm{\mathrm{k}},36} \right|^{2}$ calculated for the different $\bm{\mathrm{k}}$-vector orientations, $\theta$ and $\phi$, at $\Delta =$ 0.6, 0.8, and 1.0 meV, respectively. Phonons propagating in the angle marked in a bright color contribute to the electron-phonon interaction. Note that since the electron-phonon interaction is symmetric with respect to $z = 0$, the color-coded maps are plotted only for $0 \leq \theta \leq 0.5\pi$. (d)-(f) Color-coded maps of $\left| \Lambda_{\mu,\bm{\mathrm{k}},56} \right|^{2}$ calculated for the different $\mathbf{k}$-vector orientations, at $\Delta =$ 0.6, 0.8, and 1.0 meV, respectively.}
\label{fig3} 
\end{figure*}
We calculated $\left| \Lambda_{l,\bm{\mathrm{k}},36} \right|^{2}$ at $\left| \bm{\mathrm{k}} \right| = k_{\Delta}$. The color-coded maps of $\left| \Lambda_{l,\bm{\mathrm{k}},36} \right|^{2}$as functions of $\theta$ and $\phi$ for the $\ket{S(0,2)}$-$\ket{T_{0}(0,2)}$ energy separation $\Delta =$ 0.6, 0.8, and 1.0 meV are depicted in Fig. \ref{fig3}(a), (b), and (c), respectively. The physical constants and parameters used in the calculations are listed in Tab. \ref{Tab1}. Note that because of the symmetry of the DQD system with respect to $z = 0$, we only plot in the range of $0 \leq \theta \leq 0.5\pi$. The half inter-dot distance $d$ is set to be 96 nm. According to the color-coded maps, the electron-phonon coupling strength enhances only in $0 < \theta < 0.2\pi$. This indicates that, for $\Delta$ in this energy range, only phonons which propagate mostly perpendicularly to the $xy$-plane can interact with the electrons in the DQD. As confirmed by the definition of the electron-phonon coupling (see Eqs. \eqref{eq:gamma_0} and \eqref{eq:gamma_0_d}), only phonons whose energy is in resonance with $\Delta$ contribute to the electron-phonon coupling. In addition, the electron-phonon coupling is maximized when the dimension of the in-plane component of the phonon wavenumber vector along the quantum well surface matches the lateral dimension of the electron wavefunctions in the DQD. These two requirements are simultaneously satisfied when $\theta \sim 0.1\pi$. Furthermore, in these color-coded maps, more complicated structures originated from the inter-dot distance $d$ appear. The color-coded maps of $\left| \Lambda_{l,\bm{\mathrm{k}},56} \right|^{2}$ calculated for the same values of $\Delta$ as $\left| \Lambda_{l,\bm{\mathrm{k}},36} \right|^{2}$, are shown in Figs. \ref{fig3}(d), (e), and (f). Since $\Psi_{R^{*}}$ is expanded mostly along the $x$-axis (see APPENDIX A for the definition of $\Psi_{R^{*}}$), the larger electron-phonon coupling shows up at around $\phi = 0$ and $\pi$.

\begin{figure*}[t]
\centering
\includegraphics[width=\linewidth]{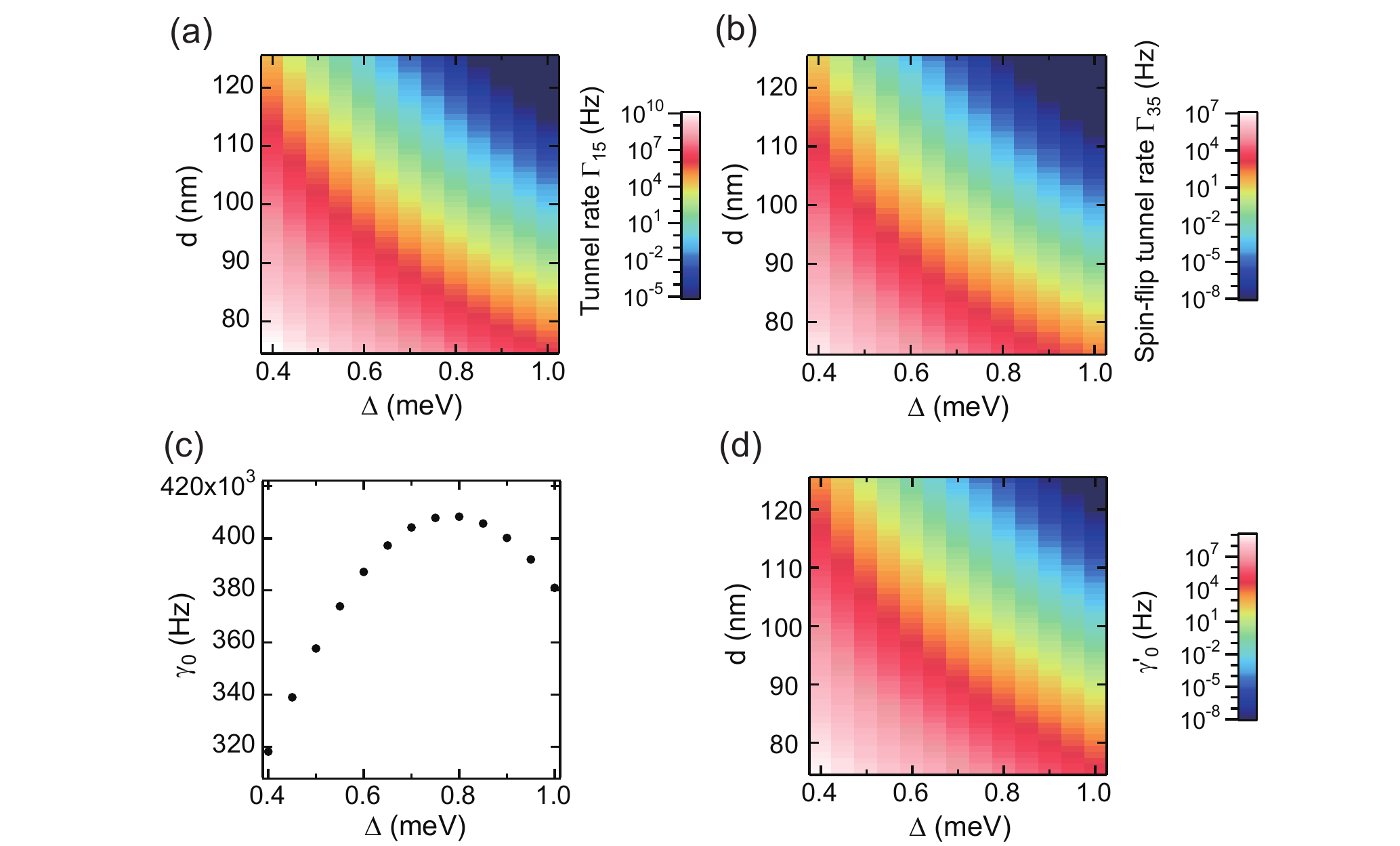} 
\caption{(a) Color-coded map of the spin-conserving tunnel rate $\Gamma_{15}$ calculated for $\Delta$ and $d$. (b) Color-coded map of the spin-flip tunnel rate $\Gamma_{35}$ calculated for $\Delta$ and $d$. (c) $\gamma_{0}$ plotted with respect to the singlet-triplet energy separation $\Delta$. (d) Color-coded map of $\gamma_{0}'$ plotted in the plane of $\Delta$ and $d$.}
\label{fig4} 
\end{figure*}

We computed the spin-conserving tunnel rate $\Gamma_{15}$ and the spin-flip tunnel rate $\Gamma_{35}$ for the ground states as functions of $\Delta$ and $d$. Note that we assume that the time scale of the inter-dot tunneling events is much smaller than the coherence time of the orbitals of electrons in the QD. In the weak coupling regime, these rates are defined by using Fermi’s golden rule \cite{HofmannPRR2020,GrabertNSSB1992,AguadoPRL1986}:
\begin{align}
\Gamma_{15}(\epsilon) & = \frac{2\pi}{\hbar}\left| \left\langle \uparrow \downarrow (1,1)\left| {\hat{\mathcal{H}}}_{\mathrm{{QD}\_{2e}}} \right|S(0,2) \right\rangle \right|^{2}P(\epsilon) \nonumber\\
 & = \frac{2\pi}{\hbar}\left| \tau_{LR} \right|^{2}P(\epsilon), \label{eq:ICtunnelRate}
\end{align}
where $\epsilon$ and $P(\epsilon)$ are the energy detuning between states 1 and 5 and the probability that the electrons exchange energy $\epsilon$ with the surrounding environment, respectively \cite{HofmannPRR2020}. $P(\epsilon)$ represents the broadening of the inter-dot tunnel rate spectrum, for example by the time independent fluctuation of the QD energy level induced by the gate voltage instability \cite{AguadoPRL1986}. We do not discuss the mechanism of $P(\epsilon)$, but from previous experimental studies, it can be assumed that $\Gamma_{15}(\epsilon) = \Gamma_{15}\exp( - \epsilon^{2}/2\delta_{\epsilon}^{2})$, where $\delta_{\epsilon}$ represents the linewidth of the inter-dot tunnel rate spectrum. Thereby, by integrating both sides of Eq. \eqref{eq:ICtunnelRate} by $\epsilon,$ the incoherent tunnel rates under the resonance condition, i.e. for $\epsilon = 0$, can be described as
\begin{align}
    \Gamma_{15} = \frac{\sqrt{2\pi}\left| \tau_{LR} \right|^{2}}{\hbar\delta_{\epsilon}}.
\end{align}
In the calculations, we adopted $\delta_{\epsilon} \sim 100\ \mu$eV estimated from our previous experiment \cite{KuroyamaPRL2022}. Similarly, the spin-flip tunnel rates between the ground states of the left and right QDs, $\Gamma_{35},$ is
\begin{align}
    \Gamma_{35} = \frac{\sqrt{2\pi}|\alpha|^{2}}{\hbar\delta_{\epsilon}} .
\end{align}
Figures. \ref{fig4}(a) and \ref{fig4}(b) show the calculated values of $\Gamma_{15}$ and $\Gamma_{35}$ plotted as functions of $\Delta$ and $d$. Both rates show a very similar dependence in the plane of $\Delta$ and $d$, meaning that the spin-flip tunnel rate of the ground states is proportional to the spin-conserving tunnel rate. This behavior is consistent to previous experimental and theoretical studies \cite{DanonPRB2013,HoffmannPRL2017,MatsuoPRR2020}. Next, we calculated the transition rates, $\gamma_{0}$ and $\gamma'_{0}$, for various values of $\Delta$ and $d$. Figure \ref{fig4}(c) depicts $\gamma_{0}$ plotted for the different $\Delta$. Note that since $\gamma_{0}$ is the transition rate of the intra-dot spin-flip process, it is independent of the inter-dot distance $d$. $\gamma_{0}$ stays at around 1 MHz for $0.4 \leq \Delta \leq 1$ meV \cite{FujisawaNature2002, SasakiPRL2005, HansonPRL2005, MeunierPRL2007}. $\gamma_{0}'$ calculated for various values of $\Delta$ and $d$ is shown in the color-coded map of Fig. \ref{fig4}(d). $\gamma_{0}'$ varies over a much wider range than $\gamma_{0}$. The dependence on $d$ is understood by the fact that the inter-dot tunnel coupling energy increases as $d$ decreases. Furthermore, the $\Delta$-dependence of $\gamma_{0}'$ is also explained by the inter-dot tunnel coupling, because $l_{0}$ decreases as $\Delta$ increases (see Eq. \eqref{eq:l_0}) in Appendix A), and $e^{- d^{2}/l_{0}^{2}}$ in $\tau_{LR}$ decreases accordingly. We chose $\Delta = 0.8$ meV and $d = 100$ nm for the following calculations, in such a way that $\Gamma_{15}$ is in the orders of a few kHz, and $\Gamma_{35}$ is below a few Hz as obtained in our previous experimental study. For those values of $\Delta$ and $d$, $\Gamma_{15} \sim 2.2$ kHz, $\Gamma_{35} \sim 0.47$ Hz, $\gamma_{0} \sim 410$ kHz, and $\gamma_{0}^{'} \sim 170$ Hz are calculated.

\begin{figure*}[t]
\centering
\includegraphics[width=\linewidth]{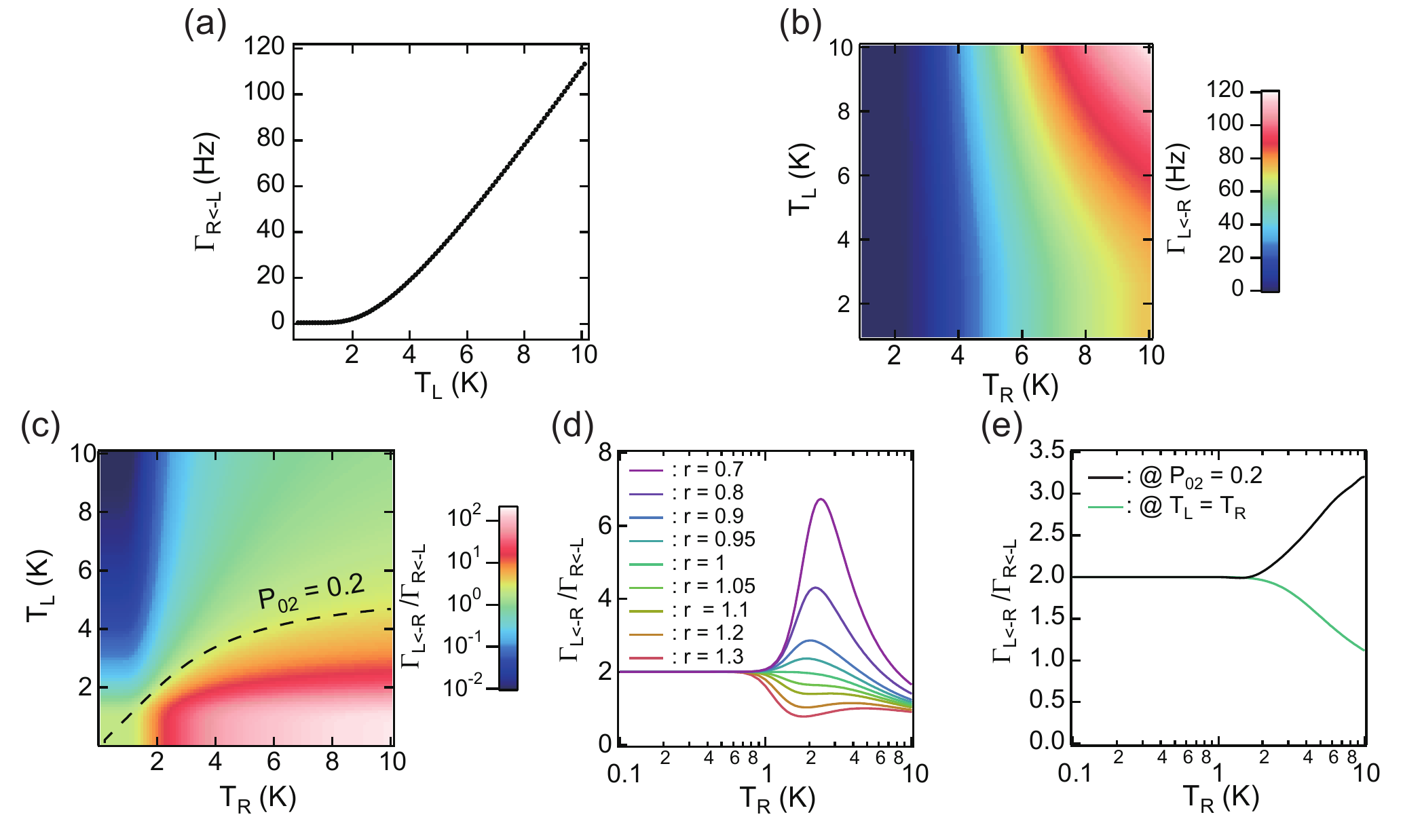} 
\caption{(a) Spin-flip tunnel rate from the left to right QD, $\Gamma_{R \leftarrow L}$, plotted as a function of $T_{L}$. (b) Color-coded map of the spin-flip tunnel rate from the right to left QD, $\Gamma_{L \leftarrow R}$, plotted as functions of $T_{L}$ and $T_{R}$. (c) Color-coded map of the spin-flip rate ratio, $\Gamma_{L \leftarrow R}/\Gamma_{R \leftarrow L}$, plotted with respect to $T_{L}$ and $T_{R}$. A dashed line indicates $T_{L}$ and $T_{R}$ satisfying $P_{02} \sim 0.2$ (see Fig. \ref{fig6}(a)). (d) $\Gamma_{L \leftarrow R}/\Gamma_{R \leftarrow L}$ ratio for various temperature gradients, i.e., $T_{L} = rT_{R}$, plotted as a function of $T_{R}$. The colors correspond to the values of $r$ as shown in the annotation in the figure. (e)$\Gamma_{L \leftarrow R}/\Gamma_{R \leftarrow L}$ ratio extracted from Fig. \ref{fig5}(c) at $P_{02} \sim 0.2$ (see a dashed line in Figs. \ref{fig5}(c) and \ref{fig6}(a)) is plotted by a black line. The same ratio at the equilibrium condition, i.e., $T_{L} = T_{R}$, is plotted by a green line, as a reference.}
\label{fig5} 
\end{figure*}
Next, we calculated the phonon-mediated spin-flip tunnel rates in the DQD when the phonon temperature is different between the left and right QDs, i.e., $T_{L}$ and $T_{R}$. If $\gamma_{0} \gg \gamma_{0}'$, we can describe the spin-flip tunnel rate from the left to right QD, $\Gamma_{R \leftarrow L}$, as follows \cite{KuroyamaPRL2022} (see Fig. \ref{fig2}(b) and Appendix C).
\begin{align}
    \Gamma_{R\leftarrow L}&\simeq\Gamma_{CB}+\Gamma_{DB}\nonumber\\
    &=\Gamma_{53}+\Gamma_{63}\nonumber\\
    &=\Gamma_{53}+\gamma_0'n_{l,k_{\Delta}}(T_L). 
    \label{eq:G_RL}
\end{align}
For the opposite tunneling direction, by considering the occupation probabilities of the spin states of (0,2), we describe the spin-flip tunnel rate from the right to left QD, $\Gamma_{L \leftarrow R}$ as
\begin{align}
    \Gamma_{L\leftarrow R}\simeq&\frac{\Gamma_{CD}}{\Gamma_{CD}+\Gamma_{DC}}\Gamma_{BC}+\frac{\Gamma_{DC}}{\Gamma_{CD}+\Gamma_{DC}}\Gamma_{BD}\nonumber\\
    =&\frac{\Gamma_{56}}{2\Gamma_{65}+\Gamma_{56}}(\Gamma_{35}+\Gamma_{45})+\frac{2\Gamma_{65}}{2\Gamma_{65}+\Gamma_{56}}\Gamma_{36}\nonumber\\
    =&2\frac{n_{l,k_{\Delta}}(T_R)+1}{3n_{l,k_{\Delta}}(T_R)+1}\Gamma_{35}\nonumber\\
    &+\frac{2n_{l,k_{\Delta}}(T_R)}{3n_{l,k_{\Delta}}(T_R)+1}\gamma_0'[1+n_{l,k_{\Delta}}(T_L)],
    \label{eq:G_LR}
\end{align}
where $\Gamma_{56}/\left( 2\Gamma_{65}+\Gamma_{56} \right)$ represents the occupation probability of either state 6 or 7, and $2\Gamma_{65}/\left( 2\Gamma_{65}+\Gamma_{56} \right)$ does that of state 5 in the $(0,2)$ charge state. We assume that the Zeeman energy induced by the $B$-field is sufficiently small compared to a thermal energy of electrons in the DQD, so that states 3 and 4, and states 6 and 7 are nearly degenerated, i.e., $\Gamma_{35} \simeq \Gamma_{45}$, $\Gamma_{53} \simeq \Gamma_{54}$, $\Gamma_{65} \simeq \Gamma_{75}$ and $\Gamma_{56} \simeq \Gamma_{57}$. We plot $\Gamma_{R \leftarrow L}$ with respect to $T_{L}$ in Fig. \ref{fig5}(a), where $d =$ 96 nm and $\Delta =$ 0.8 meV. When $T_{L} > 2$ K, since the population of phonons with the energy $\Delta$ starts to increase, the phonon-mediated inter-dot transitions from state 3 to 6 and from state 4 to 7 start to take place frequently, and $\Gamma_{R \leftarrow L}$ increases with increasing $T_{L}$. Similarly, $\Gamma_{L \leftarrow R}$ is plotted as functions of $T_{L}$ and $T_{R}$ in the color-coded map of Fig. \ref{fig5}(b) and increases above about $T_{R} = 2$ K due to the frequent phonon-mediated spin-flip excitations from state 5 to states 6 and 7.

A more intriguing feature appears in the temperature dependence of the ratio of $\Gamma_{L \leftarrow R}$/ $\Gamma_{R \leftarrow L}$. We plot it with respect to $T_{L}$ and $T_{R}$ in the color-coded map depicted in Fig. \ref{fig5}(c). The map can be divided into five characteristic regions: When $T_{L},$ $T_{R} < 2$ K, the ratio is constant at 2, which is simply determined by the ratio of $\Gamma_{BC}$ to $\Gamma_{CB}$. When $T_{L} < 2$ K and $T_{R} > 2$ K, the phonon excitation processes take place only in the right QD. Therefore, only $\Gamma_{L \leftarrow R}$ increases, and the ratio $\Gamma_{L \leftarrow R}$/ $\Gamma_{R \leftarrow L}$ increases accordingly. In contrast, when $T_{L} > 2$ K and $T_{R} < 2$ K, the phonon excitation processes take place only in the left QD. Therefore, only $\Gamma_{R \leftarrow L}$ increases, and the ratio decreases accordingly. For $T_{R} > T_{L} > 2$ K, the ratio is larger than that in the equilibrium condition of $T_{L} = T_{R}$. This behavior can be understood by the fact that the phonon excitation in the right QD is more frequent than that in the left QD. For $T_{L} > T_{R} > 2$ K, the behavior is reversed. To show the imbalance of the spin-flip tunnel rates more clearly, we plot the $\Gamma_{L \leftarrow R}$ /$\Gamma_{R \leftarrow L}$ vs $T_{R}\ $for different temperature gradients, i.e., $T_{L} = rT_{R}$, in Fig. \ref{fig5}(d). First, for the equilibrium case of $r = 1$, the ratio stays at 2 for low temperatures and decreases above around 1 K. Note that the factor 2 is originated from the ratio of $\left( \Gamma_{35} + \Gamma_{45} \right)/\Gamma_{53}$. Furthermore, the ratio $\Gamma_{L \leftarrow R}/\Gamma_{R \leftarrow L}$ approaches 2/3 at higher $T_{L}$. This is explained by a fact that states 5, 6, and 7 are equally populated at higher $T_{L}$ and $T_{R}$ and that $\gamma_{0}^{'} \gg \Gamma_{35},\Gamma_{53}$. Thus, the value of 2/3 is determined by the proportion of occupation probability of either state 6 or 7 out of states 5, 6, and 7. For $T_{L} < T_{R}$, the ratio is clearly different from the behavior of the equilibrium case and increases at $T_{L} >$ 1K. The increase of the ratio is larger for the smaller $r$, i.e., larger temperature gradient between the QDs. For $T_{L} > T_{R}$, the ratio can be smaller than 2/3 when the temperature gradient is larger. 

\begin{figure*}[t]
\centering
\includegraphics[width=\linewidth]{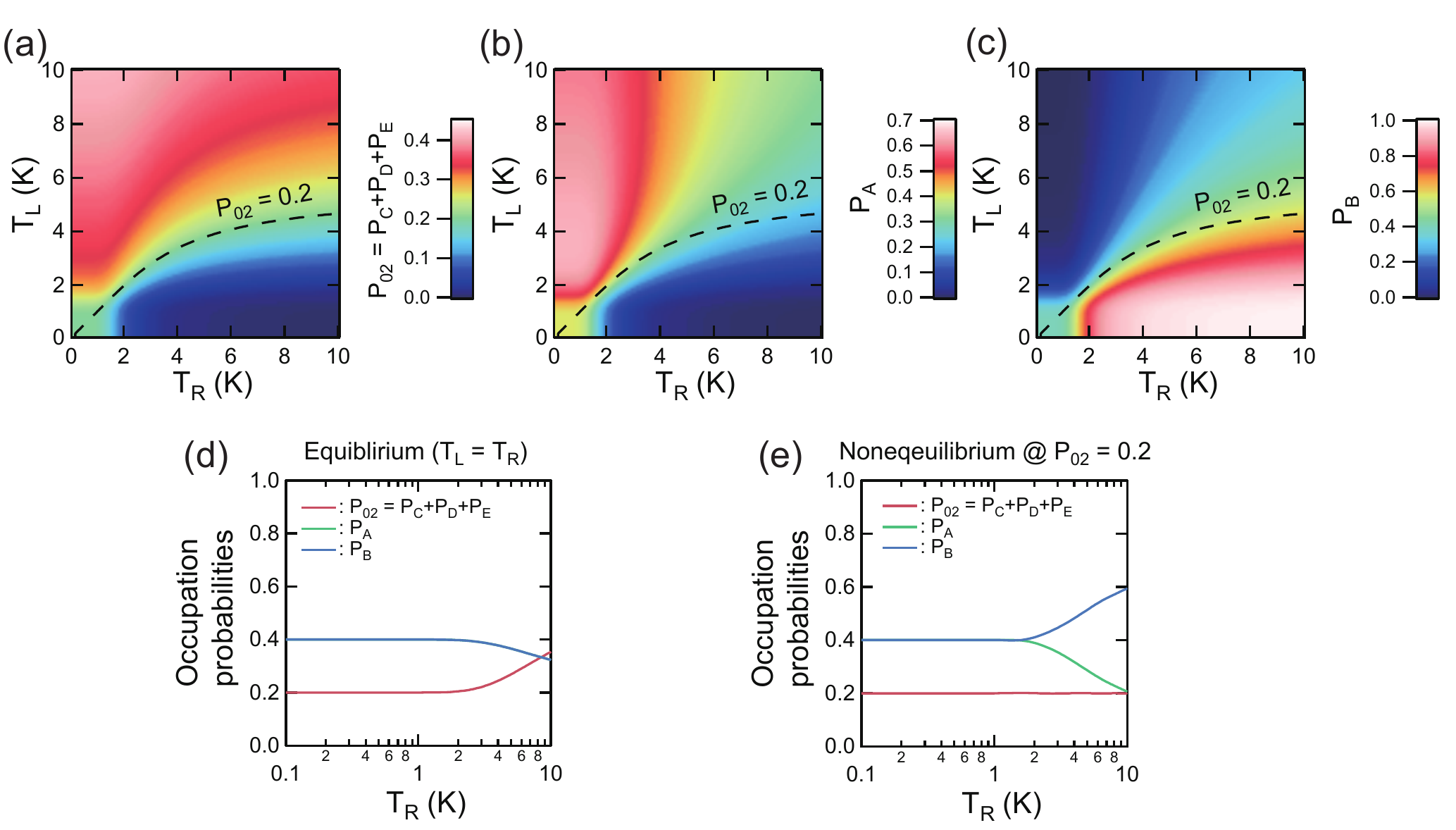} 
\caption{(a)-(c) Color-coded maps of $P_{02}$, $P_{A}$, and $P_{B}$, respectively, calculated for $T_{L}$ and $T_{R}$. (d) $P_{02}$ (red), $P_{A}$ (green), and $P_{B}$ (blue) plotted as a function of $T_{R}$ under the equilibrium condition, i.e., $T_{L} = T_{R}$. $P_{A}$, and $P_{B}$ take the same value. (e) $P_{A}$ and $P_{B}$ extracted from Figs. \ref{fig6}(b) and (c), respectively, at $P_{02} \sim 0.2$ (see a dashed line in Figs. \ref{fig6}(a-c)). }
\label{fig6} 
\end{figure*}

\section{Occupation probabilities of the two-electron spin states}
Next, we calculated the occupation probabilities from state A to E, (refer to the transition diagram in Fig. \ref{fig2}(b)) for different values of $T_{L}$ and $T_{R}$. We considered the occupation probabilities of $|S(0,2)\rangle$, $|T_{\pm}(0,2)\rangle$, and $|T_{0}(0,2)\rangle$ in the $(0,2)$ charge state, $P_{C}$, $P_{D}$, and $P_{E}$, and those of the anti-parallel spin states $P_{A}$, and the parallel spin states, $P_{B}$, in the $(1,1)$ charge state. Figures \ref{fig6}(a)-(c) show the color-coded maps of $P_{C} + P_{D} + P_{E} \equiv P_{02}$, $P_{A}$, and $P_{B}$ , respectively. When $T_{L},T_{R} < 2$ K, since the occupations of the ground states dominate, $P_{02} = 0.2$, and $P_{A},\ P_{B} = 0.4$. In the case of $T_{R} > T_{L} > 2$ K, $P_{A}$ decreases, and $P_{B}$ increases. This behavior is explained by more phonon excitation from state C to D and the imbalance between $\Gamma_{BD}$ and $\Gamma_{DB}$ (i.e., $\Gamma_{BD} > \Gamma_{DB}$). Thus, state B is more populated by the phonon-mediated spin-flip transition from state C to B via state D. On the other hand, in the case of $T_{L} > T_{R} > 2$ K, since more phonon excitation from state B to D takes place, and state B is depopulated, resulting in the increase of the occupation probability of state A, $P_{A}$, and of the probability of the $(0,2)$ charge state, $P_{02}$. Next, we considered the equilibrium case, i.e., $T_{L} = T_{R}$, and the occupation probabilities $P_{02}$, $P_{A}$, and $P_{B}\ $are shown in Fig. \ref{fig6}(d). Note that $P_{A}$ and $P_{B}$ overlap completely in the equilibrium condition. Above $T_{R} = 2$ K, $P_{A}$ and $P_{B}$ start to decrease from 0.4 while $P_{02}$ increases from 0.2. 

According to our experimental results (see Fig. 4(b) in Ref. \cite{KuroyamaPRL2022}), $P_{02}$ was constant at around 0.2, while $P_{A}$ decreased and $P_{B}$ increased from 0.4. These behaviors of the occupation probabilities are clearly different from those calculated in the equilibrium condition shown in Fig. \ref{fig6}(d). In order to compare the theoretically calculated occupation probabilities with the experimental result, we extract the $P_{A}$ and $P_{B}$ for $T_{L}$ and $T_{R}$ that satisfy $P_{02} \sim 0.2$ (see a dashed line in Fig. \ref{fig6}(a)). Figure \ref{fig6}(e) shows $P_{02}$, $P_{A}$, and $P_{B}$ along the dashed line. Unlike the equilibrium case shown in Fig. \ref{fig6}(d), while $P_{A}$ decreases above $T_{R} =$ 2 K, and $P_{B}$ increases around the same temperature. These behaviors of $P_{A}$ and $P_{B}$ are in very good agreement with our experimental result. We can conclude that the characteristic imbalance of the occupation probability between parallel and anti-parallel spin configurations shown in Ref. \cite{KuroyamaPRL2022} is created not by the phonon temperature increase but the phonon temperature gradient in the DQD system. In addition, the black dashed line of $P_{02}\left( T_{L},T_{R} \right) = 0.2$ in Fig. \ref{fig6}(a) predicts what kind of phonon temperature gradient was actually generated by a nearby phonon source QD \cite{KuroyamaPRL2022}.

Finally, we make a brief comment on the ratio of $\Gamma_{L\leftarrow R}$ to $\Gamma_{R\leftarrow L}$. As mentioned in the previous paragraph, our previous experimental study showed $P_{02} (T_L,T_R)\sim0.2$ for all values of a bias voltage applied on a nearby phonon source QD. Therefore, we calculated the ratio as a function of $T_R$ under the condition of $P_{02} (T_L,T_R)\sim0.2$, as depicted by a black line in Fig. \ref{fig5}(e). A green line in Fig. \ref{fig5}(e) is the ratio calculated at the equilibrium condition, i.e., $T_L=T_R$, extracted from Fig. \ref{fig5}(d) and is shown as a reference data. The ratio $\Gamma_{L\leftarrow R}/\Gamma_{R\leftarrow L}$ calculated under the nonequilibrium condition increases above $T_R=2$ K, while that of the equilibrium condition decreases. Although in our previous study \cite{KuroyamaPRL2022}, we plotted the ratio as a function of the applied bias voltage of a nearby phonon source QD instead of the phonon temperatures, this behavior of  $\Gamma_{L\leftarrow R}/\Gamma_{R\leftarrow L}$ is in good agreement with our experimental result as well (see Fig. S5 in Supplemental Material of Ref. \cite{KuroyamaPRL2022}).

\section{Conclusion}
In conclusion, we theoretically investigated the phonon-mediated spin-flip tunnel processes in the two-electron DQD under a phonon temperature gradient over the DQD, to reproduce our previous experimental result \cite{KuroyamaPRL2022}. The spin-flip tunnel processes via the $\ket{T_{\pm}(0,2)}$ states are strongly enhanced when the thermal energy of acoustic phonons exceeds the excitation energy of the triplet states. Furthermore, the spin-flip tunnel rates in opposite tunneling directions to each other is significantly affected by the phonon temperature gradient. The occupation probabilities of the respective spin-states are significantly modified as well by the phonon temperature gradient. These behaviors are in excellent agreement with our previous experimental result. In addition, the calculations regarding the occupation probabilities indicates that as long as $P_{02} (T_L,T_R)$ is kept at a certain constant value like our previous experiment, the DQD in the nonequilibrium environment realizes spin pumping-like effect makes either the  parallel spin state or anti-parallel spin state in the $(1,1)$ charge state more occupied. Since the spin-pumping-like effect can be regarded as thermodynamic work conducting on spins by a heat flow at the DQD, our DQD system can work as mesoscopic heat engines using single spins as working substances. If the phonon temperatures in QDs can be controlled individually in the coupled QD systems, more interesting spin-related thermodynamic functions of QDs would be realized.

\section{Acknowledgement}
This work was supported by Grant-in-Aid for Scientific Research (S) (No. JP19H05610), JST CREST (No. JPMJCR15N2 and JPMJCR1675), JST Moonshot R$\&$D grant (No. JPMJMS226B), MEXT Quantum Leap Flagship Program grant (No. JPMXS0118069228), Grant-in-Aid for Young Scientists (No. JP20K14384), and Grant-in-Aid for JSPS Fellows (No. JP16J03037, No. JP19J01737).  Y. T. is supported by Grant-in-Aid for Scientific Research (C) (No. JP18K03479) and JST Moonshot R$\&$D–MILLENNIA Program (Grant No. JPMJMS2061).

\section*{Appendix A: Electron wavefunctions in the DQD}
We assume that the Fock-Darwin states are suitable for representing the electron wavefunctions of the DQD\cite{FockZP1928, DarwinCUP1931, BarnesPRB2011,NielsenPRB2013}. The ground states in the left QD, $\psi_L$, those in the right QD, $\psi_R$, and the first excited state in the right QD, $\psi_{R^*}$ are
\begin{align}
\psi_L(\bm{\mathrm{r}})&=\frac{1}{\sqrt{\pi}l_0}\exp\left[-\frac{(x+d)^2+y^2}{2l_0^2}\right],\\
\psi_R(\bm{\mathrm{r}})&=\frac{1}{\sqrt{\pi}l_0}\exp\left[-\frac{(x-d)^2+y^2}{2l_0^2}\right],\\
\psi_{R^*}(\bm{\mathrm{r}})&=\psi^+_{R^*}(\bm{\mathrm{r}})\cos\phi_a+\psi^-_{R^*}(\bm{\mathrm{r}})\mathrm{e^{i\phi_b}}\sin\phi_a, 
\end{align}
respectively. For, $\psi^{\pm}_{R^*}(\bm{\mathrm{r}})$, we use the linear combination of the wavefunctions of the 2p orbitals:
\begin{align}
\psi^{\pm}_{R^*}(\bm{\mathrm{r}})&=\frac{x-d\pm iy}{\sqrt{\pi}l_0^2}\exp\left[-\frac{(x-d)^2+y^2}{2l_0^2}\right].
\end{align}
Considering the shape of the DQD potential in the sample used in Ref. \cite{KuroyamaPRL2022}, the wavefunction along the $x$-axis has lower energy. Therefore, the wavefunctions of the excited states are assumed to expand along the $x$ axis, i.e., $\phi_a\sim\pi/4$, and $\phi_b\sim0$. We chose $\phi_a=\pi/4$ and $\phi_b=0.1\pi$ in such a way that $\gamma_0'$, $\Gamma_{15}$, and $\Gamma_{35}$ are in the same order as the values obtained in our previous study \cite{KuroyamaPRL2022}.
The Fock-Darwin wavefunctions of the different dots overlap each other, whereas the wavefunctions in the QDs should be orthogonal to each other. To avoid the overlaps, we introduced the extended Hund-Mulliken approach \cite{WhitePRB2018}. The wavefunctions having a small inter-dot overlap are mostly orthogonalized by the following treatments.
\begin{align}
\Psi_L(\bm{\mathrm{r}})&\equiv\frac{1}{N_g}\left[\psi_L(\bm{\mathrm{r}})-\frac{S}{2}\psi_R(\bm{\mathrm{r}})-2S_p\psi_{R^*}(\bm{\mathrm{r}})\right], \label{eq:Psi_l}\\
\Psi_R(\bm{\mathrm{r}})&\equiv\frac{1}{N_g}\left[\psi_R(\bm{\mathrm{r}})-\frac{S}{2}\psi_L(\bm{\mathrm{r}})\right],\label{eq:Psi_r}\\
\Psi_{R^*}(\bm{\mathrm{r}})&\equiv\frac{1}{N_e}\left[\psi_{R^*}(\bm{\mathrm{r}})+S_p\psi_L(\bm{\mathrm{r}})\right]\label{eq:Psi_re},
\end{align}
where $S$ and $S_p$ are defined as
\begin{align}
S&\equiv\int d^2\bm{\mathrm{r}}\psi_L^*(\bm{\mathrm{r}})\psi_R(\bm{\mathrm{r}}),\\
S_p&\equiv\int d^2\bm{\mathrm{r}}\psi_L^*(\bm{\mathrm{r}})\psi_{R^*}(\bm{\mathrm{r}}),
\end{align}
and $N_{g} \simeq 1 - 3S^{2}/4$ and $N_{e} \simeq 1 + S_{p}^{2}$ are the normalization coefficients. We note that these newly defined wavefunctions, $\Psi_{L}$, $\Psi_{R}$, and $\Psi_{{R}^{*}}$, are almost orthonormalized in the limit of the weak inter-dot coupling. For the $z$ component of the wavefunctions, $\psi_{z}(z)$, the lowest eigenstate of the rectangular well model with infinite barriers is considered with the quantum well length $l_{z}$. The wavefunction is described as $\psi_{z}(z) = \sqrt{2/l_{z}}\cos\left( \pi z/l_{z} \right)$, where $- l_{z}/2 \leq z \leq l_{z}/2$. The $z$-component of the wavefunctions, $\psi_{z}$, is multiplied to the wavefunctions of the $xy$-plane, Eqs. \eqref{eq:Psi_l}-\eqref{eq:Psi_re}, for the calculation of the electron-phonon interaction, because $\psi_{z}$ involves the momentum matching between an electron and a phonon in the electron-phonon interaction. For simplicity, the wavefunction $\psi_{z}$ is omitted in the following expressions.

The eigenenergies of the ground states in the left and right QDs, $\epsilon_L$ and $\epsilon_R$, and that of the first excited state in the right QD, $\epsilon_{R^*}$, are calculated by using $\hat{\mathcal{H}}_0$ as
\begin{align}
\int d^2\bm{\mathrm{r}}\Psi_L^*(\bm{\mathrm{r}})\hat{\mathcal{H}}_0(\bm{\mathrm{r}})\Psi_L(\bm{\mathrm{r}})&\equiv\epsilon_L,\\
\int d^2\bm{\mathrm{r}}\Psi_R^*(\bm{\mathrm{r}})\hat{\mathcal{H}}_0(\bm{\mathrm{r}})\Psi_R(\bm{\mathrm{r}})&\equiv\epsilon_R,\\
\int d^2\bm{\mathrm{r}}\Psi_{R^*}^*(\bm{\mathrm{r}})\hat{\mathcal{H}}_0(\bm{\mathrm{r}})\Psi_{R^*}(\bm{\mathrm{r}})&\equiv\epsilon_{R^*}.
\end{align}
The off-diagonal matrix elements of $\hat{\mathcal{H}}_0$ represent the tunnel coupling between the ground states in the two dots, $\tau_{LR}$, and that between the ground state in the left QD and the first-excited state in the right QD, $\tau_{LR^*}$:
\begin{align}
\int d^2\bm{\mathrm{r}}\Psi_L^*(\bm{\mathrm{r}})\hat{\mathcal{H}}_0(\bm{\mathrm{r}})\Psi_R(\bm{\mathrm{r}})&\equiv\tau_{LR},\\
\int d^2\bm{\mathrm{r}}\Psi_L^*(\bm{\mathrm{r}})\hat{\mathcal{H}}_0(\bm{\mathrm{r}})\Psi_{R^*}(\bm{\mathrm{r}})&\equiv\tau_{LR^*},
\end{align}
where the parameters $\tau_{LR}$, and $\tau_{LR^*}$ are complex numbers in general.

Furthermore, the energy separation, $\Delta$, between $\ket{S(0,2)}$ and $\ket{T_{0}(0,2)}$ is determined by the summation of the orbital spacing $\hbar\omega_{0}$ in the QD and the Coulomb interaction between two electrons \cite{GolovachPRB2008}. By using the ratio of $\Delta$ to the QD orbital spacing $\hbar\omega_{0}$, $\delta_{ST}$, the electron confinement length in the QD, $l_{0}$, is determined by $\Delta$ and $\delta_{ST}$:
\begin{align}
l_{0} = \sqrt{\frac{\hbar}{m^{*}\omega_{0}}} = \hbar\sqrt{\frac{\delta_{ST}}{m^{*}\Delta}}\label{eq:l_0}
\end{align}
According to Ref. \cite{GolovachPRB2008}, the value of $\delta_{ST}$ is determined by the confinement length of an electron in the QD $l_{0}$ and the Bohr radius defined by the Coulomb interaction. For the calculations in this paper, we adopted the typical reported value of the ratio $\delta_{ST} \sim 0.5$ \cite{HansonPRB2004}.

\section*{APPENDIX B: Spin-orbit interactions}
For the Dresselhaus-type spin-orbit interaction, the momentum terms in the $xy$-plane are
\begin{align}
{\hat{p}}_{x'}{\hat{\sigma}}_{x'} + {\hat{p}}_{y'}{\hat{\sigma}}_{y'} = & - {\hat{p}}_{x}{\hat{\sigma}}_{x}\cos 2\delta + {\hat{p}}_{x}{\hat{\sigma}}_{y}\sin 2\delta \nonumber\\
& + {\hat{p}}_{y}{\hat{\sigma}}_{x}\sin 2\delta + {\hat{p}}_{y}{\hat{\sigma}}_{y}\cos 2\delta.
\end{align}
The Rashba-type spin-orbit interaction is transformed to the expression in the $xy$-plane:
\begin{align}
\hat{p}_x'\hat{\sigma}_y'-\hat{p}_y'\hat{\sigma}_x'=&
\hat{p}_{x}\hat{\sigma}_{y}-\hat{p}_{y}\hat{\sigma}_{x}.
\end{align}
Thus,
\begin{align}
    {\hat{\mathcal{H}}}_{\mathrm{so}} = \frac{\hbar}{m^{*}}\left[ - \frac{\cos 2\delta}{\lambda_{\mathrm{D}}}{\hat{p}}_{x}{\hat{\sigma}}_{x} + \left( \frac{\sin 2\delta}{\lambda_{\mathrm{D}}} + \frac{1}{\lambda_{\mathrm{R}}} \right){\hat{p}}_{x}{\hat{\sigma}}_{y} \right.\nonumber\\
\left( \frac{\sin 2\delta}{\lambda_{\mathrm{D}}} - \frac{1}{\lambda_{\mathrm{R}}} \right){\hat{p}}_{y}{\hat{\sigma}}_{x}\left. \  + \frac{\cos 2\delta}{\lambda_{\mathrm{D}}}{\hat{p}}_{y}{\hat{\sigma}}_{y} \right].
\end{align}
Because the $B$-field is applied along the $- y$ axis, the spin is oriented to the same direction. Therefore, the terms including ${\hat{\sigma}}_{y}$ do not contribute to the spin-flip process and can be dismissed. Summing up the transformations, the spin-orbit interaction Hamiltonian in the $xy$-plane is described as
\begin{align}
{\hat{\mathcal{H}}}_{\mathrm{so}} =& \frac{\hbar}{m^{*}}\left[ - \frac{\cos 2\delta}{\lambda_{\mathrm{D}}}{\hat{p}}_{x}{\hat{\sigma}}_{x} \right.\nonumber\\
&+\left. \left( \frac{\sin 2\delta}{\lambda_{\mathrm{D}}} - \frac{1}{\lambda_{\mathrm{R}}} \right){\hat{p}}_{y}{\hat{\sigma}}_{x} \right]. 
\end{align}
Furthermore, since the $-y$ axis is the quantized axis of a spin, $n_z^{\mathrm{spin}}$, in the spin-space, the Pauli matrix $\hat{\sigma}_{x'}$ is replaced with $-\hat{\sigma}_{y}$, and the Hamiltonian described as Eq. \eqref{eq:SOI}.

\section*{APPENDIX C: Occupation probabilities}
To evaluate the occupation probabilities from spin state A to E in Fig. \ref{fig2}(b), we formed the following Master equations using the transition rates:
\begin{align}
\frac{dP_{A}}{dt}  =& - \left( \Gamma_{CA} + \Gamma_{EA} \right)P_{A}(t) + \Gamma_{AC}P_{C}(t)\nonumber\\
&+ \Gamma_{AE}P_{E}(t), \\
\frac{dP_{B}}{dt} =& - \left( \Gamma_{CB} + \Gamma_{DB} \right)P_{B}(t) + \Gamma_{BC}P_{C}(t)\nonumber\\
&+ \Gamma_{BD}P_{D}(t), \\
\frac{dP_{C}}{dt} =& \Gamma_{CA}P_{A}(t) + \Gamma_{CB}P_{B}(t)\nonumber\\
&- \left( \Gamma_{AC} + \Gamma_{BC} + \Gamma_{DC} \right)P_{C}(t) + \Gamma_{CD}P_{D}(t), \\
\frac{dP_{D}}{dt} =& \Gamma_{DB}P_{B}(t) + \Gamma_{DC}P_{C}(t) \nonumber\\
&- \left( \Gamma_{BD} + \Gamma_{CD} \right)P_{D}(t), \\
\frac{dP_{E}}{dt} =& \Gamma_{EA}P_{A}(t) - \Gamma_{AE}P_{E}(t), 
\end{align}
where $P_{i}(t)$ $(i = A,\cdots,E)$ is the occupation probability of state $i$ in the transition diagram of Fig. \ref{fig2}(b). We solved these equations for a case of the steady states, i.e., $dP_{i}/dt = 0$, and evaluated the occupation probabilities of all spin states. The transition rates are calculated by the following relations: $\Gamma_{AC} = \Gamma_{15} +\Gamma_{25}=2\Gamma_{15}$, $\Gamma_{CA} = \Gamma_{51} = \Gamma_{52}$, $\Gamma_{BC} = \Gamma_{35} + \Gamma_{45}=2\Gamma_{35}$, $\Gamma_{CB} = \Gamma_{53} = \Gamma_{54}$, $\Gamma_{CD} = \Gamma_{56} = \Gamma_{57}$, $\Gamma_{DC} = 2\Gamma_{65} = 2\Gamma_{75}$, $\Gamma_{BD} = \Gamma_{36} = \Gamma_{47}$, $\Gamma_{DB} = \Gamma_{63} = \Gamma_{74}$, $\Gamma_{AE} = 2\Gamma_{18} = 2\Gamma_{28} = \Gamma_{36}$, and $\Gamma_{EA} = \Gamma_{81} = \Gamma_{82} = \Gamma_{63}/2$.

%

\end{document}